\documentclass[prb,twocolumn,preprintnumbers,amsmath,amssymb]{revtex4}

\usepackage{graphicx}
\graphicspath{{fig/}}
\usepackage{dcolumn}
\usepackage{bm}
\usepackage{amsmath,amssymb}
\usepackage[T1]{fontenc}
\usepackage{textcomp}
\usepackage{color}

\begin{document}

\title{Universal $1/3$-suppression of magnonic shot noise in diffusive insulating magnets 
}

\author{Kouki Nakata,$^{1}$ Yuichi Ohnuma,$^{2}$ and Mamoru Matsuo$^{2}$}

\affiliation{$^1$Advanced Science Research Center, Japan Atomic Energy Agency, Tokai, Ibaraki 319-1195, Japan  \\
$^2$Kavli Institute for Theoretical Sciences, University of Chinese Academy of Sciences, Beijing 100190, China
}

\date{\today}

\begin{abstract}
Extending a Boltzmann-Langevin theory to magnons, we show a universality of current-noise suppression in diffusive systems
against the difference of quantum-statistical properties of bosons and fermions.
To this end, starting from a quantum kinetic equation for magnons subjected to thermal gradient in dilute impurities,
we derive a bosonic counterpart of the semiclassical Boltzmann-Langevin equation for electrons
and evaluate a magnonic current-noise in the diffusive insulating magnet.
We theoretically discover that compared with a Poissonian shot noise of magnons in an insulating ferromagnetic junction,
the magnonic shot noise is suppressed in the diffusive insulating bulk magnet
and noise-to-current ratio (Fano factor) at low temperatures exhibits a universal behavior,
i.e., the same suppression $1/3$ as the one for electron transport in diffusive conductors,
despite the difference of quantum-statistical properties.
Finally, we show that our predictions are within experimental reach with current device and measurement technologies.
\end{abstract}

\maketitle

\section{Introduction}
\label{sec:Intro}

Since the observation of quasiequilibrium Bose-Einstein condensation of magnons \cite{demokritov} in an insulating ferromagnet
and the associated transport phenomena \cite{MagnonSupercurrent},
many magnonic counterparts of electron transport have been established both theoretically and experimentally over the last decade
(e.g., the Josephson effect \cite{KKPD} and the Wiedemann-Franz law \cite{magnonWF} of magnons, etc.; 
see Ref. [\onlinecite{ReviewMagnon}] for details).
For further development of magnon-based spintronics, dubbed magnonics \cite{MagnonSpintronics}, aimed at utilizing magnons as a carrier of information in units of the Bohr magneton $\mu_{\rm{B}}$,
magnonic current-noise, recently measured in Ref. [\onlinecite{magnonNoiseMeasurement}],
has been attracting much attention as the most promising new direction
since, as seen in a saying of Landauer \cite{LandauerNoise} ``The noise is the signal'', nonequilibrium noise at low temperatures (i.e., shot noise) provides abundant information about transport beyond conductance. 
The well-known example is the suppression of Fano factor (noise-to-current ratio) in diffusive conductors \cite{NoiseRev}.

The $1/3$-suppression of electron shot noise in metallic diffusive systems due to dilute-impurity scattering,
which highlights the fact that charge transport in a bulk is 
different from a junction \cite{DLnoise3},
has been established theoretically by several different approaches \cite{NoiseRev}.
Beenakker and B$\ddot {\rm{u}}$ttiker \cite{Beenakker1/3} have demonstrated it by using a Landauer approach (i.e., a scattering matrix approach),
while Nagaev \cite{Nagaev1/3} by a semiclassical Boltzmann-Langevin approach;
making use of the semiclassical approach, Sukhorukov and Loss \cite{DLnoise6} have found that phase coherence, 
which is a priori included into the scattering matrix approach,
is not relevant to the suppression of shot noise.
Those theoretical predictions, the $1/3$-suppression of electron shot noise in diffusive conductors, 
has been experimentally confirmed \cite{1/3experiment}.
However, still the intriguing question remains an open issue;
whether the Pauli exclusion principle of electrons is essential or not for the noise suppression,
and moreover, whether the noise suppression factor `$1/3$' is robust or not against the difference of the quantum-statistical properties of bosons and fermions.

In this paper we provide a solution to this fundamental challenge by developing the Boltzmann-Langevin approach \cite{Nagaev1/3}
into magnons in a diffusive insulating magnet.
Magnons, low-energy magnetic excitations, are chargeless bosonic quasiparticles and therefore, in contrast to electrons,
they are free from the Pauli exclusion principle.
Thereby considering magnons in a diffusive insulating bulk magnet due to dilute impurity scattering
while carefully taking into account the difference from electrons that magnons do not possess a sharp Fermi surface associated with the Fermi energy,
we shed new light on the question. 

This paper is organized as follows.
In Sec. \ref{sec:bulk} starting from a quantum kinetic equation for magnons,
we derive a Boltzmann-Langevin theory for magnons and evaluate a magnonic spin current-noise in a diffusive insulating magnet.
We then find a universality of current-noise suppression in diffusive systems 
against the difference of quantum-statistical properties of bosons and fermions at low temperatures.
In Sec. \ref{sec:Observation} we provide some insight into experiments.
Finally, we summarize in Sec. \ref{sec:summary} and remark several issues in Sec. \ref{sec:discussion}.
Technical details are deferred to the Appendix.

\section{Magnonic spin current-noise in diffusive bulk}
\label{sec:bulk}

\subsection{Boltzmann-Langevin equation for magnons}
\label{subsec:Boltzmann}

We consider a three-dimensional insulating bulk magnet being subjected to thermal gradient along the $x$-axis
where magnons in a uniform dc magnetic field along the $z$-axis weakly interact with dilute impurities.
The magnon dispersion is $ \hbar \omega _{\mathbf{k}} := \epsilon _{\mathbf{k}}  + \Delta $,
where the magnon gap $ \Delta  $ is characterized by the applied magnetic field and the easy-axis spin anisotropy, etc.
The quantum kinetic equation \cite{haug} of magnons is given by
$ (\partial _t + {\mathbf{v}} \cdot  \partial _{{\mathbf{r}}}) G^{<}_{{\mathbf{k}}, \omega , {\mathbf{r}}, t}
 = (1/{i\hbar }) (G^{<}_{{\mathbf{k}}, \omega , {\mathbf{r}}, t} \Sigma ^{>}_{{\mathbf{k}}, \omega , {\mathbf{r}}, t}
- G^{>}_{{\mathbf{k}}, \omega , {\mathbf{r}}, t} \Sigma ^{<}_{{\mathbf{k}}, \omega , {\mathbf{r}}, t})$,
where $ {\mathbf{v}} = \partial _{{\mathbf{k}}} \omega _{\mathbf{k}}$ is the magnon velocity,
and $ G^{<(>)}$ and $ \Sigma ^{<(>)} $ are the lesser (greater) components of 
the bosonic nonequilibrium Green's function \cite{Schwinger,Schwinger2,Keldysh,haug,mahan,tatara,kita}
and the self-energy, respectively.
Within the Born approximation \cite{haug}, the self-energy due to magnon-dilute impurity scattering becomes
$  \Sigma _{{\mathbf{k}}, \omega , {\mathbf{r}}, t} = [{V}/{(2\pi)^3}] \int d {\mathbf{k}}' 
G_{{\mathbf{k}}', \omega , {\mathbf{r}}, t} \mid    V_{{\mathbf{k}}, {\mathbf{k}}'}   \mid ^2 $,
where  $V \equiv L^3$ is the volume of the system and 
$  V_{{\mathbf{k}}, {\mathbf{k}}'}   $  is the impurity potential.
The quantum kinetic equation of magnons thus becomes
$ (\partial _t + {\mathbf{v}} \cdot  \partial _{{\mathbf{r}}}) G^{<}_{{\mathbf{p}}, \omega , {\mathbf{r}}, t}
 = (1/{i\hbar }) [{V}/{(2\pi \hbar )^3}] \int d {\mathbf{p}}'  \mid    V_{{\mathbf{p}}, {\mathbf{p}}'}   \mid ^2
(G^{<}_{{\mathbf{p}}, \omega , {\mathbf{r}}, t} G^{>}_{{\mathbf{p}}', \omega , {\mathbf{r}}, t}
- G^{>}_{{\mathbf{p}}, \omega , {\mathbf{r}}, t} G ^{<}_{{\mathbf{p}}', \omega , {\mathbf{r}}, t})$,
where $ {\mathbf{p}} = \hbar   {\mathbf{k}}$.

Under the Kadanoff-Baym ansatz \cite{haug,KBtextbook} 
each bosonic nonequilibrium Green's function is characterized by the nonequilibrium Bose-distribution function 
$f_{{\mathbf{p}}, \omega , {\mathbf{r}}, t}$,
$ G^{<}_{{\mathbf{p}}, \omega , {\mathbf{r}}, t} 
= 2 i {\rm{Im}}  G^{\rm{r}}_{{\mathbf{p}}, \omega , {\mathbf{r}}, t} f_{{\mathbf{p}}, \omega , {\mathbf{r}}, t} $,
$ G^{>}_{{\mathbf{p}}, \omega , {\mathbf{r}}, t} 
= 2 i {\rm{Im}}  G^{\rm{r}}_{{\mathbf{p}}, \omega , {\mathbf{r}}, t} (1+f_{{\mathbf{p}}, \omega , {\mathbf{r}}, t}) $,
and within the quasiparticle approximation \cite{haug}
the retarded Green's function $G^{\rm{r}}$ becomes
$  {\rm{Im}}  G^{\rm{r}} = - \pi \delta (\hbar \omega - \hbar  \omega _{\mathbf{p}})  $.
Thereby integrating the quantum kinetic equation over the frequency $\omega $, $ \int d \omega $, 
it reduces to the semiclassical \cite{mahan} Boltzmann equation for magnons,
\begin{eqnarray}
 (\partial _t + {\mathbf{v}} \cdot  \partial _{{\mathbf{r}}}) f_{{\mathbf{p}}, {\mathbf{r}}, t} = {\cal{I}}[J],
 \label{eqn:Eq1} 
\end{eqnarray}
where  
the magnon-impurity collision integral ${\cal{I}}[J] $ is characterized by the particle currents $J({\mathbf{p}}, {\mathbf{p}}') $ 
from the state $ {\mathbf{p}}$ to  $ {\mathbf{p}}'$
(for simplicity we omit the subscript  ${\mathbf{r}}$ and $ t$, etc. when it is unnecessary; e.g., $  f_{{\mathbf{p}}, {\mathbf{r}}, t}  \equiv  f_{\mathbf{p}}$);
\begin{subequations}
\begin{eqnarray}
  {\cal{I}}[J] &=& \frac{V}{(2\pi \hbar )^3} \int d {\mathbf{p}}' [J({\mathbf{p}}', {\mathbf{p}}) -  J({\mathbf{p}}, {\mathbf{p}}')], 
 \label{eqn:Eq2a}      \\
  J({\mathbf{p}}, {\mathbf{p}}') &=&  W({\mathbf{p}}, {\mathbf{p}}') f_{{\mathbf{p}}} (1+f_{{\mathbf{p}}'}),  
\label{eqn:Eq2b}    \\
  W({\mathbf{p}}, {\mathbf{p}}') &=&  - \frac{2}{\hbar } \mid V_{{\mathbf{p}}, {\mathbf{p}}'} \mid ^2
  {\rm{Im}} G^{\rm{r}}_{{\mathbf{p}}', \omega , {\mathbf{r}}, t} \mid _{\omega =\omega _{\mathbf{p}}}.
  \label{eqn:Eq2c}
\end{eqnarray}
\end{subequations}
Since $ V_{{\mathbf{p}}, {\mathbf{p}}'} = V_{{\mathbf{p}}', {\mathbf{p}}} $ for a spatially symmetric potential,
the probability $W({\mathbf{p}}, {\mathbf{p}}') $ of scattering per unit time due to the impurity potential is symmetric 
$ W({\mathbf{p}}, {\mathbf{p}}') =  W({\mathbf{p}}', {\mathbf{p}})  $
in terms of  ${\mathbf{p}}$ and ${\mathbf{p}}'$.
Integrating the semiclassical Boltzmann equation [Eq. (\ref{eqn:Eq1})]
over  $  {\mathbf{k}}$, $ \int d {\mathbf{k}}$, 
it reduces to the continuity equation of magnons,
$ \partial _t \rho ({\mathbf{r}}, t) + {\mathbf{\nabla }} \cdot   {\mathbf{j}}  = 0$,
where 
$  \rho ({\mathbf{r}}, t) :=   \int d {\mathbf{k}} f_{{\mathbf{k}}, {\mathbf{r}}, t} $
and
$   {\mathbf{j}} :=   \int d {\mathbf{k}} {\mathbf{v}} f_{{\mathbf{k}}, {\mathbf{r}}, t}  $.
Throughout this paper, we thus work under the assumption that the number of magnons is conserved.

Due to the randomness of the scattering process, the particle currents $J({\mathbf{p}}, {\mathbf{p}}') $  between the states 
$ {\mathbf{p}}$ and $ {\mathbf{p}}'$ fluctuate \cite{Kogan}
and therefore it can be written by 
$    J({\mathbf{p}}, {\mathbf{p}}') =  \bar {J}({\mathbf{p}}, {\mathbf{p}}') + \delta J({\mathbf{p}}, {\mathbf{p}}')  $
and
$   f_{{\mathbf{p}}} = {\bar {f}}_{{\mathbf{p}}} + \delta f_{{\mathbf{p}}}  $ accordingly,
where $\bar {J} ({\bar {f}})$ is the average current (distribution) and $\delta J (\delta f)$ represents the fluctuations.
Thus following the approach of Refs. [\onlinecite{Nagaev1/3,Kogan}]
and thus taking the effects of the fluctuations into the Boltzmann equation [Eq. (\ref{eqn:Eq1})], 
we obtain the Boltzmann-Langevin equation for magnons;
\begin{eqnarray}
 (\partial _t + {\mathbf{v}} \cdot  \partial _{{\mathbf{r}}}) ({\bar {f}}_{{\mathbf{p}}} + \delta f_{{\mathbf{p}}}) 
=  {\cal{I}}[\bar {J} + \delta J].
\label{eqn:Eq3}
\end{eqnarray}
Since $  {\cal{I}}[\bar {J} + \delta J]= {\cal{I}}[\bar {J}] + {\cal{I}}[\delta J]   $
the fluctuating Langevin source \cite{NoiseRev} $\xi _{{\mathbf{p}}} := {\cal{I}}[\delta J]  $,
being zero on the statistical average  $  \langle  \xi _{{\mathbf{p}}}  \rangle =0  $,
becomes 
\begin{eqnarray}
\xi _{{\mathbf{p}}} = \frac{V}{(2\pi \hbar )^3} \int d {\mathbf{p}}' 
[\delta J({\mathbf{p}}', {\mathbf{p}}) -  \delta J({\mathbf{p}}, {\mathbf{p}}')],
\label{eqn:Eq4}
\end{eqnarray}
and assuming a Poisson process given by
$
\langle \delta J({\mathbf{p}}_1, {\mathbf{p}}_2, {\mathbf{r}}, t)  
\delta J({\mathbf{p}}'_1, {\mathbf{p}}'_2, {\mathbf{r}}', t') \rangle 
= [(2 \pi \hbar)^6/V] \delta ({\mathbf{p}}_1 - {\mathbf{p}}'_1) \delta ({\mathbf{p}}_2 - {\mathbf{p}}'_2)
\delta ({\mathbf{r}} - {\mathbf{r}}')  \delta (t-t') {\bar {J}} ({\mathbf{p}}_1, {\mathbf{p}}_2, {\mathbf{r}}, t)
$,
the statistical average of the correlations between the Langevin sources become \cite{NoiseRev,WhiteNoise}
$ 
\langle  \xi ({\mathbf{r}}, {\mathbf{p}}, t)   \xi ({\mathbf{r}}', {\mathbf{p}}', t')    \rangle   
= V \delta ({\mathbf{r}} - {\mathbf{r}}')  \delta (t-t')
\{ 
\delta ({\mathbf{p}} - {\mathbf{p}}') \int   d {\mathbf{p}}'' 
[{\bar {J}} ({\mathbf{p}}'', {\mathbf{p}}, {\mathbf{r}}, t) + {\bar {J}} ({\mathbf{p}}, {\mathbf{p}}'', {\mathbf{r}}, t)]
- [{\bar {J}} ({\mathbf{p}}, {\mathbf{p}}', {\mathbf{r}}, t) + {\bar {J}} ({\mathbf{p}}', {\mathbf{p}}, {\mathbf{r}}, t)]
\}
$.

We remark that the above semiclassical Boltzmann-Langevin theory [Eqs. (\ref{eqn:Eq3}) and (\ref{eqn:Eq4})]
is identified with the magnonic counterpart of the theory for electrons developed by Nagaev \cite{Nagaev1/3};
in the case of electrons (i.e., fermions),  the particle current of Eq. (\ref{eqn:Eq2b}) is replaced by
$J_{({\rm{F}})} \propto { {f}}_{{\mathbf{p}} ({\rm{F}})}  (1 - { {f}}_{{\mathbf{p}}' ({\rm{F}})})   $
instead of 
$J_{({\rm{F}})}  \propto { {f}}_{{\mathbf{p}} ({\rm{F}})}  (1 + { {f}}_{{\mathbf{p}}' ({\rm{F}})})   $,
where ${ {f}}_{\mathbf{p} ({\rm{F}})}$ is the nonequilibrium Fermi-distribution function characterized by the Fermi energy.
The sign inversion between magnons and electrons in the particle current
can be traced back to the greater component of the nonequilibrium Green's functions \cite{kita}
 in the quantum kinetic equation \cite{haug};
$    G^{>}_{\mathbf{p}} \propto   (1 + { {f}}_{\mathbf{p}})   $ for bosons, 
while
$    G^{>}_{{\mathbf{p}}({\rm{F}})} \propto   (1 - { {f}}_{{\mathbf{p}}({\rm{F}})})   $ for fermions.
Those result from the difference of the quantum-statistical properties that 
bosons are free from the Pauli exclusion principle imposed for fermions.

\subsection{Magnonic shot noise}
\label{subsec:MagnonicShotNoise}

Assuming that the system is slightly out of equilibrium,
the relaxation-time approximation \cite{mahan} provides 
$  {\cal{I}}[\bar {J} + \delta J] = - ({f_{\mathbf{p}} -f_{\rm{eq.}}})/{\tau _{\mathbf{p}}}   $,
where 
$ f_{\rm{eq.}}$ is the equilibrium configuration and
$ {\tau _{\mathbf{p}}}$ represents the relaxation time.
Since $  {\cal{I}}[\bar {J} + \delta J]= {\cal{I}}[\bar {J}] + {\cal{I}}[\delta J]   $,
the Boltzmann-Langevin equation for magnons [Eq. (\ref{eqn:Eq3})] provides 
\begin{subequations}
\begin{eqnarray}
 {\cal{I}}[\bar {J}]  
&=& (\partial _t + {\mathbf{v}} \cdot  \partial _{{\mathbf{r}}}) {\bar {f}}_{{\mathbf{p}}}   
= - \frac{\bar {f}_{\mathbf{p}} - f_{\rm{eq.}}}{\tau _{\mathbf{p}}},   
 \label{eqn:Eq5a}  \\
{\cal{I}}[\delta J]   
&=& \xi _{{\mathbf{p}}}  =  - \frac{\delta  {f}_{\mathbf{p}} }{\tau _{\mathbf{p}}}. 
\label{eqn:Eq5b}
\end{eqnarray}
\end{subequations}
Thus the fluctuations of the nonequilibrium Bose-distribution function $ \delta  {f}_{\mathbf{p}}$
is characterized by the fluctuating Langevin source $  \xi _{{\mathbf{p}}} $ 
and the relaxation time [Eq. (\ref{eqn:Eq5b})].
From Eq. (\ref{eqn:Eq5a})
the relaxation time $ {\tau _{\mathbf{k}}}$ is given by 
$  1/{\tau _{\mathbf{k}}} = \Sigma _{{\mathbf{k}}'} W({\mathbf{k}}, {\mathbf{k}}')
(1- {\mathbf{v}}_{\mathbf{k}}\cdot {\mathbf{v}}_{{\mathbf{k}}'}/\mid  {\mathbf{v}}_{\mathbf{k}}  \mid ^2) $.

Since thermal gradient is applied along the $x$-axis,
the resulting magnonic spin current is given by 
$I_x = g \mu_{\rm{B}} L^2 \int [{d {\mathbf{k}}}/{(2\pi)^3}] v_{k_x} f_{\mathbf{k}}$,
where $g$ is the $g$-factor of the constituent spins.
The average and the fluctuations of the magnonic spin current, $\bar {I}_x$ and $ \delta  {I}_x $, respectively, become
\begin{subequations}
\begin{eqnarray}
\bar {I}_x &=& g \mu_{\rm{B}} L^2 \int  \frac{d {\mathbf{k}}}{(2\pi)^3} v_{k_x} {\bar {f}}_{\mathbf{k}}, 
\label{eqn:Eq6a}  \\
\delta  {I}_x &=& g \mu_{\rm{B}} L^2 \int  \frac{d {\mathbf{k}}}{(2\pi)^3} v_{k_x}  \delta f_{\mathbf{k}}.
\label{eqn:Eq6b}
\end{eqnarray}
\end{subequations}
Assuming a steady state in terms of time $ {\bar {f}}_{{\mathbf{k}}, {\mathbf{r}}, t} =  {\bar {f}}_{{\mathbf{k}}}(x) $,
Eq. (\ref{eqn:Eq5a}) provides
$  {\bar {f}}_{{\mathbf{k}}}(x)  = - \tau _{\mathbf{k}} v_{k_x} \partial _x  {\bar {f}}_{{\mathbf{k}}} + f_{\rm{eq.}}$,
and from Eq. (\ref{eqn:Eq6a}) we obtain 
$ \bar {I}_x  =   - g \mu_{\rm{B}} L^2 \int  [{d {\mathbf{k}}}/{(2\pi)^3}] 
{\tau _{\mathbf{k}}}  (v_{k_x})^2  \partial _x  {\bar {f}}_{\mathbf{k}} $.
Under the boundary conditions \cite{Nagaev1/3}
$  {\bar {f}}_{{\mathbf{k}}}(x=-L/2) \equiv  f_{\rm{L}}    $ and $  {\bar {f}}_{{\mathbf{k}}}(x= L/2) \equiv  f_{\rm{R}}    $
where the temperature is $T_{\rm{L}}$ and $T_{\rm{R}}$, respectively,
the spatial average $  \langle \bar {I}_x  \rangle  := (1/L) \int_{-L/2}^{L/2} dx  \bar {I}_x  $ becomes
\begin{eqnarray}
 \langle \bar {I}_x  \rangle  =   g \mu_{\rm{B}} L \int  \frac{d {\mathbf{k}}}{(2\pi)^3} {\tau _{\mathbf{k}}}  (v_{k_x})^2  
(f_{\rm{L}}-f_{\rm{R}}).
\label{eqn:Eq7}
\end{eqnarray}

Taking both the statistical average and the spatial average,
we introduce the correlations between the fluctuations of the magnonic spin current as
\begin{eqnarray}
 {\cal{S}}(t, t') :=  \langle \langle \delta  {I}_x (t)  \delta  {I}_x (t')  \rangle \rangle.
 \label{eqn:Eq8}
\end{eqnarray}
From Eq. (\ref{eqn:Eq6b})
it becomes
$   {\cal{S}}(t, t') = (g \mu_{\rm{B}})^2  L^4   \int  [{d {\mathbf{k}}}/{(2\pi)^3}]   \int  [{d {\mathbf{k}}'}/{(2\pi)^3}]  
v_{k_x} v_{k'_x}  \langle  \langle \delta  {f}_{\mathbf{k}}({\mathbf{r}}, t) \delta  {f}_{{\mathbf{k}}'}({\mathbf{r}}', t')   \rangle \rangle
$,
where within the relaxation-time approximation [Eq. (\ref{eqn:Eq5b})]
$ \langle  \langle \delta  {f}_{\mathbf{k}}({\mathbf{r}}, t) \delta  {f}_{{\mathbf{k}}'}({\mathbf{r}}', t')   \rangle \rangle
= {\tau _{\mathbf{k}}} {\tau _{{\mathbf{k}}'}} 
\langle  \langle   \xi _{{\mathbf{k}}} ({\mathbf{r}}, t)   \xi _{{\mathbf{k}}'} ({\mathbf{r}}', t')  \rangle \rangle
$.
Assuming that the impurity potential is localized in space,
the Fourier component becomes momentum-independent \cite{haug}
 $ \mid   V_{{\mathbf{k}}, {\mathbf{k}}'}   \mid  ^2 = u^2 n_{\rm{imp.}}  $,
  where the impurity concentration $n_{\rm{imp.}}$ is dilute.
Thereby a straightforward calculation provides \cite{SeeSuppl,WhiteNoise}
\begin{eqnarray}
 {\cal{S}}(t, t')
&=& 2 (g \mu_{\rm{B}})^2  L  \delta (t-t')   \int  \frac{d {\mathbf{k}}}{(2\pi)^3}    {\tau _{\mathbf{k}}}  (v_{k_x})^2   \nonumber    \\
&\cdot &     \langle {\bar {f}}_{\mathbf{k}}(x)  [1 + {\bar {f}}_{\mathbf{k}}(x)] \rangle.
\label{eqn:Eq9}
\end{eqnarray}
From the effective kinetic equation of magnons in the diffusive regime
the spatial average becomes \cite{SeeSuppl}
\begin{eqnarray}
    \langle {\bar {f}}_{\mathbf{k}}(x)  [1 + {\bar {f}}_{\mathbf{k}}(x)] \rangle
&=& \frac{f_{\rm{R}}-f_{\rm{L}}}{6} 
{\rm{coth}} \Big(\frac{\beta _{\rm{L}} \hbar \omega _{\mathbf{k}}}{2} 
- \frac{\beta _{\rm{R}} \hbar \omega _{\mathbf{k}}}{2} \Big)   \nonumber \\
&-& \frac{1}{3}  \Big( k_{\rm{B}} T_{\rm{L}} \frac{\partial  f_{\rm{L}}(\epsilon )}{\partial \epsilon} 
+ k_{\rm{B}} T_{\rm{R}} \frac{\partial  f_{\rm{R}}(\epsilon )}{\partial \epsilon} \Big),
  \label{eqn:Eq10}      \      \         \        \         \          \       
\end{eqnarray}
where
$  f_{\rm{L(R)}}(\epsilon ) := ({\rm{e}}^{\beta _{\rm{L(R)}} \epsilon }-1)^{-1} $,
$\epsilon :=  \hbar \omega _{\mathbf{k}} $,
and
$ \beta _{\rm{L(R)}} :=  (k_{\rm{B}} T_{\rm{L(R)}})^{-1}$.
The first term of the right-hand side in Eq. (\ref{eqn:Eq10}) represents the nonequilibrium noise 
and the second term is the thermal noise (i.e., equilibrium noise) \cite{ThermalNoise,ThermalNoise2}.

At low temperatures 
\begin{eqnarray}
 k_{\rm{B}} T_{\rm{L(R)}} \ll  \Delta ,
 \label{eqn:Eq11}
\end{eqnarray}
thermal noise is suppressed and becomes negligibly small,
 \begin{eqnarray}
    \langle {\bar {f}}_{\mathbf{k}}(x)  [1 + {\bar {f}}_{\mathbf{k}}(x)] \rangle
\stackrel{\rightarrow }{=}  \frac{f_{\rm{L}}-f_{\rm{R}}}{6} 
{\rm{coth}} \Big(\frac{\beta _{\rm{R}} \hbar \omega _{\mathbf{k}}}{2} 
- \frac{\beta _{\rm{L}} \hbar \omega _{\mathbf{k}}}{2} \Big),    
  \label{eqn:Eq12}     \      \      \     
\end{eqnarray}
and 
$  {\rm{coth}} ({\beta _{\rm{R}} \hbar \omega _{\mathbf{k}}}/{2} 
- {\beta _{\rm{L}} \hbar \omega _{\mathbf{k}}}/{2})  \stackrel{\rightarrow }{=}   1 $
when
$   {\beta _{\rm{R}} \hbar \omega _{\mathbf{k}}}/{2} - {\beta _{\rm{L}} \hbar \omega _{\mathbf{k}}}/{2}   \gg  1 $;
i.e.,
 \begin{eqnarray}
 \frac{T_{\rm{L}}  -  T_{\rm{R}}}{T_{\rm{R}}}   \gg   \frac{2 k_{\rm{B}}  T_{\rm{L}}}{\Delta}.
 \label{eqn:Eq13}
 \end{eqnarray}
Finally, in the magnonic shot noise regime 
$  ({T_{\rm{L}}  -  T_{\rm{R}}})/{T_{\rm{R}}}   \gg   {2 k_{\rm{B}}  T_{\rm{L}}}/{\Delta} $
at low temperatures $k_{\rm{B}} T_{\rm{L(R)}} \ll  \Delta$,
the correlations $ {\cal{S}}(t, t') =  {\cal{S}}(t - t') $ between the fluctuations of the magnonic spin current reduce to \cite{WhiteNoise}
\begin{eqnarray}
{\cal{S}}(t, t')
\stackrel{\rightarrow }{=} \frac{1}{3} (g \mu_{\rm{B}})^2  L  \delta (t-t')   
 \int  \frac{d {\mathbf{k}}}{(2\pi)^3}    {\tau _{\mathbf{k}}}  (v_{k_x})^2   (f_{\rm{L}}-f_{\rm{R}}).    
 \label{eqn:Eq14}    \         \        
\end{eqnarray}
Defining the magnonic spin current-noise spectrum by 
$ {\cal{S}}(\Omega ) := \int d \tilde{t}  {\rm{e}}^{i \Omega  \tilde{t} }  {\cal{S}}(\tilde{t})   $ 
with $ \tilde{t} := t-t' $
and focusing on the dc-limit $\Omega =0$,
the Fano factor for magnons ${\cal{F}} $ becomes
\begin{eqnarray}
{\cal{F}} :=  \frac{ {\cal{S}}(\Omega =0 )}{g \mu_{\rm{B}}  \langle \bar {I}_x  \rangle }   \stackrel{\rightarrow }{=} \frac{1}{3}.
\label{eqn:Eq15}
\end{eqnarray}
In contrast to a Poissonian shot noise of magnons (i.e., ${\cal{F}} = 1$) in an insulating ferromagnetic junction \cite{KNmagnonNoiseJunction},
the magnonic shot noise is thus suppressed to $1/3$ in a diffusive insulating bulk magnet.
This is the main result of this paper.

\subsection{Universality of the $1/3$-noise suppression}
\label{subsec:univerdsality}

The above result [Eq. (\ref{eqn:Eq15})],
 the $1/3$-suppression of the Fano factor for magnons in diffusive insulating bulk magnets due to dilute impurity scattering,
is robust against details of the magnon dispersion.
Therefore any magnons with a gapped energy spectrum following the Boltzmann-Langevin equation [Eq. (\ref{eqn:Eq3})]
exhibit the $1/3$-noise suppression, 
the same suppression as the one for electron transport in diffusive conductors \cite{Beenakker1/3,Nagaev1/3,DLnoise6,NoiseRev}.
This means that the Pauli exclusion principle is not essential for the noise suppression;
quite remarkably, even bosons at low temperatures exhibit 
the same $1/3$-noise suppression as the one for fermions 
in spite of the fact that the quantum-statistical properties of bosons and fermions are
fundamentally different, in particular in the low temperature region where quantum effects dominate.
This highlights the universality of the $1/3$-noise suppression in diffusive systems due to dilute impurity scattering;
the noise suppression factor `$1/3$' is {\textit{universal}} despite the difference of the quantum-statistical properties of bosons and fermions.

We stress that even within the present semiclassical approach,
quantum-statistical properties of bosons have been included into the theory
via the nonequilibrium Bose-distribution function ${\bar {f}}_{\mathbf{k}}$
and the current-noise 
$ {\cal{S}}(t, t') \propto   \langle {\bar {f}}_{\mathbf{k}}  (1 + {\bar {f}}_{\mathbf{k}}) \rangle $ [Eq. (\ref{eqn:Eq9})];
in the case of fermions \cite{Nagaev1/3}, the distribution function is replaced by the fermionic counterpart ${\bar {f}}_{\mathbf{k} ({\rm{F}})}$ characterized by the Fermi energy and the form of the current-noise is modified into
$ {\cal{S}}_{({\rm{F}})}(t, t')  \propto    \langle {\bar {f}}_{{\mathbf{k}} ({\rm{F}})}  (1 - {\bar {f}}_{{\mathbf{k}} ({\rm{F}})}) \rangle   $
instead of 
$ {\cal{S}}_{({\rm{F}})}(t, t')  \propto    \langle {\bar {f}}_{{\mathbf{k}} ({\rm{F}})}  (1 + {\bar {f}}_{{\mathbf{k}} ({\rm{F}})}) \rangle   $.
The sign inversion between bosons and fermions in the current-noise
can be traced back to the greater component of the nonequilibrium Green's functions \cite{kita} in the quantum kinetic equation;
$    G^{>}_{\mathbf{k}} \propto   (1 + {\bar {f}}_{\mathbf{k}})   $ for bosons, while
$    G^{>}_{{\mathbf{k}}({\rm{F}})} \propto   (1 - {\bar {f}}_{{\mathbf{k}}({\rm{F}})})   $ for fermions.
Those result from the difference of the quantum-statistical properties that 
bosons are free from the Pauli exclusion principle imposed for fermions.

Lastly, we remark that microscopically taking into account the above-mentioned detailed difference between bosons and fermions,
we have found in Fano factor the universality;
while the current-noise itself is different from each other $ {\cal{S}}(t, t')   \not=  {\cal{S}}_{({\rm{F}})}(t, t')  $
due to the difference of the Bose- and Fermi-distribution function,
we can extract the universal behavior by taking noise-to-current ratio.

\section{Estimates for experiments}
\label{sec:Observation}

Recently, measurement of a magnonic current-noise in an insulating ferromagnet has been reported in Ref. [\onlinecite{magnonNoiseMeasurement}]
and is now within experimental reach with current device and measurement techniques.
Thereby using the measurement scheme, it would be interesting to test our theoretical predictions experimentally.
For an estimate, we assume the following experiment parameter values to realize the condition 
for the $1/3$-suppression of magnonic shot noise [Eqs. (\ref{eqn:Eq11}) and (\ref{eqn:Eq13})];
$ \Delta = 0.2$meV,
$ T_{\rm{L}} -  T_{\rm{R}}   =1$mK,
$T_{\rm{R}}=10$mK.
In the low temperature regime \cite{magnon10mK,tabuchi,tabuchiScience}, 
interaction effects (e.g., magnon-magnon and magnon-phonon interactions)
can be assumed to become negligibly small \cite{adachiphonon,Tmagnonphonon}.
Moreover, the measurement scheme \cite{magnonNoiseMeasurement}
ensures that the background noise is suppressed and negligibly small, i.e., the level of the background noise is at least an order of magnitude smaller than that measured magnonic current-noise.
Given these estimates,
we conclude that the observation of our theoretical predictions, while being challenging, 
seems within experimental reach with current device and measurement technologies.


\section{Summary}
\label{sec:summary}

Assuming the magnon gap
and making use of the semiclassical Boltzmann-Langevin equation for magnons 
while
carefully taking into account the difference from electrons,
we have demonstrated the universality of the $1/3$-noise suppression in diffusive systems
that it is robust against the difference of quantum statistical properties of bosons and fermions.
Regardless of details of the magnon dispersion, noise-to-current ratio (Fano factor) of magnons subjected to thermal gradient in dilute impurities reduces to $1/3$ in the magnonic shot noise regime at low temperatures;
the temperature scale for magnons is given by the magnon gap $\Delta/k_{\rm{B}} $,
while it is the Fermi temperature for electrons.
Quite remarkably, in spite of the fact that the quantum-statistical properties of bosons and fermions are
fundamentally different, in particular in the low temperature regime where quantum effects dominate,
magnons at low temperatures exhibit the same $1/3$-noise suppression as the one for electrons;
even the suppression factor `$1/3$' is universal despite quantum-statistical properties of bosons and fermions.

Following the approach by Nagaev \cite{Nagaev1/3} and extending the semiclassical theory to magnons,
we have discovered the universality.
While our approach is thus straightforward, we believe our finding newly exploiting spin current-noise suppression in an insulating `bulk' magnet
\footnote{As to spin current-noise in `junctions', see, e.g., Refs. [\onlinecite{KNmagnonNoiseJunction,noise2018,Kamra2,Kamra3}].
}
is non-trivial.
Thus providing a new direction for development of noise-suppressed spintronics devices,
our work serves as an intersection among magnonics, mesoscopic physics, and quantum optics.


\section{Discussion}
\label{sec:discussion}

To conclude a few comments on our approach are in order. 
First, 
magnonic shot noise [Eqs. (\ref{eqn:Eq11}) and (\ref{eqn:Eq13})] 
is produced in a magnon system with a gapped energy spectrum ($\Delta \not=0$),
while the shot noise itself is expected not to be realized in a gapless magnon system ($\Delta =0$)
because thermal noise is not suppressed due to the absence of the magnon gap 
and thereby it becomes relevant 
\footnote{
Noise consists of two parts \cite{KNmagnonNoiseJunction}; thermal noise (i.e., equilibrium noise) due to thermal fluctuations and 
nonequilibrium noise driven by an external or a thermal force.
When thermal noise is suppressed and nonequilibrium noise becomes dominant, the noise is identified with magnonic shot noise.
}.
Since the main purpose of this paper is to provide a solution to those fundamental questions that whether the Pauli exclusion principle of fermions is essential or not to the $1/3$-suppression of shot noise and whether the noise suppression factor is robust or not against the difference of the quantum-statistical properties, 
starting from the quantum kinetic equation for magnons with a gapped energy spectrum
in the dilute impurities
and following the approach by Nagaev \cite{Nagaev1/3},
we have developed the bosonic counterpart of the semiclassical Boltzmann-Langevin theory for electrons.
It is expected that interaction effects (e.g., magnon-magnon and magnon-phonon interactions)
can be taken into account in the same way as the one for electrons \cite{NoiseRev}, 
while it remains an open issue and deserves further study.
Quantum-mechanical corrections (e.g., vertex correction-induced weak localization) to the noise spectrum 
can be evaluated by relaxing some of the approximations made in this paper, e.g., Born approximation,
and it will be addressed elsewhere in the near future
following Ref. [\onlinecite{Altshuler}].

Second,
this paper has provided quantitative understanding of the universality of $1/3$ shot noise suppression.
We believe an approach following the work by Dorokhov \cite{Dorokhov,kamenev}
or the one from the viewpoint of 
decoherence/entanglement \cite{TextbookQuantumOptics,TextbookQuantumOptics2,DLnoise4}
will be one of the most promising strategies to obtain qualitative understanding,
which deserves further study.
Note that since the standard scattering matrix approach \cite{kamenev,Beenakker1/3} a priori assumes
the Pauli exclusion principle and a sharp Fermi surface associated with the Fermi energy,
it requires careful analysis \cite{NoiseRev} 
if one tries to apply the Landauer formula based on the scattering matrix theory
to bosonic cases.

Last,
from the universality of the shot noise suppression found in this paper, 
one might suspect that photons, i.e., bosons with a linear dispersion $ \omega _{\mathbf{k}}  \propto  k$,
also exhibit the same noise suppression.
However, in contrast to the magnons with a gapped energy spectrum studied in this paper,
photons are intrinsically gapless.
Moreover, on top of the fluctuations due to the randomness of impurity scattering process \cite{Nagaev1/3,Kogan}
described by Eq. (\ref{eqn:Eq3}),
vacuum fluctuations of electromagnetic fields \cite{GlauberQuantumOptics,ReviewPhotonics} 
provide a new source of fluctuations to photonic systems; 
the source term associated with the vacuum fluctuations 
(e.g., random absorption and emission)
is newly added into the Boltzmann-Langevin equation for photons \cite{BeenakkerPhoton}.
We thus conclude that the super-Poissonian photonic noise predicted theoretically in Ref. [\onlinecite{BeenakkerPhoton}]
results from vacuum fluctuations
or the gapless energy spectrum or both
instead of their bosonic nature \footnote{
Following quantum optics \cite{GlauberQuantumOptics,TextbookQuantumOptics,TextbookQuantumOptics2} 
see Ref. [\onlinecite{PhotonNoiseBunching}] for the distinction between (sub-) super-Poissonian noise and (anti-) bunching. 
See also Refs. [\onlinecite{TextbookQuantumOptics,TextbookQuantumOptics2}] for photon-bunching and -antibunching,
and Ref. [\onlinecite{BunchingAntibunching}] for the application to electron systems.
}
.

Note that in this paper we have studied the spin current-noise in a strongly spin exchange-coupled region (i.e., bulk magnet),
while Refs. [\onlinecite{KNmagnonNoiseJunction,noise2018,Kamra2,Kamra3}] have investigated the one in a weakly spin exchange-coupled region (i.e., across the junction interface); perturbative calculations in terms of the spin exchange interaction are applicable to the latter, while not to the former.



\begin{acknowledgments}
This work (KN) is supported by Leading Initiative for Excellent Young Researchers, MEXT, Japan.
We would like to thank M. Oka, T. Kato, and H. Chudo for helpful discussion.
\end{acknowledgments}

\appendix

\section{Correlations between fluctuations of magnon currents}
\label{sec:SM1}

In this Appendix, we provide details on the derivation of magnonic spin current-noise in the diffusive regime.


In the main text, taking both the statistical average and the spatial average 
we have defined the correlations between the fluctuations of the magnonic spin current subjected to thermal gradient along the $x$-axis by
$ {\cal{S}}(t, t') :=  \langle \langle \delta  {I}_x (t)  \delta  {I}_x (t')  \rangle \rangle   $,
which becomes
\begin{eqnarray}
{\cal{S}}(t, t') &=& (g \mu_{\rm{B}})^2  L^4   
\int  \frac{{d {\mathbf{k}}}}{{(2\pi)^3}}   \int  \frac{{d {\mathbf{k}}'}}{{(2\pi)^3}}  
v_{k_x} v_{k'_x}     \nonumber   \\
 &\times  &   \langle  \langle \delta  {f}_{\mathbf{k}}({\mathbf{r}}, t) \delta  {f}_{{\mathbf{k}}'}({\mathbf{r}}', t')   \rangle \rangle .
 \label{eqn:SMEq1}
\end{eqnarray}
Within the relaxation-time approximation,
the fluctuations of the nonequilibrium Bose-distribution function $ \delta  {f}_{\mathbf{k}}$
is characterized by the fluctuating Langevin source $  \xi _{{\mathbf{k}}} $  
being zero on the statistical average  $  \langle  \xi _{{\mathbf{k}}}  \rangle =0  $,
and it becomes
$     {\delta  {f}_{\mathbf{k}} }  = - {\tau _{\mathbf{k}}}  \xi _{{\mathbf{k}}}  $
and
$ \langle  \langle \delta  {f}_{\mathbf{k}}({\mathbf{r}}, t) \delta  {f}_{{\mathbf{k}}'}({\mathbf{r}}', t')   \rangle \rangle
= {\tau _{\mathbf{k}}} {\tau _{{\mathbf{k}}'}} 
\langle  \langle   \xi _{{\mathbf{k}}} ({\mathbf{r}}, t)   \xi _{{\mathbf{k}}'} ({\mathbf{r}}', t')  \rangle \rangle
$
accordingly.
Since the Poisson process assumed in the main text provides
$ 
\langle  \xi ({\mathbf{r}}, {\mathbf{p}}, t)   \xi ({\mathbf{r}}', {\mathbf{p}}', t')    \rangle   
= V \delta ({\mathbf{r}} - {\mathbf{r}}')  \delta (t-t')
\{ 
\delta ({\mathbf{p}} - {\mathbf{p}}') \int   d {\mathbf{p}}'' 
[{\bar {J}} ({\mathbf{p}}'', {\mathbf{p}}, {\mathbf{r}}, t) + {\bar {J}} ({\mathbf{p}}, {\mathbf{p}}'', {\mathbf{r}}, t)]
- [{\bar {J}} ({\mathbf{p}}, {\mathbf{p}}', {\mathbf{r}}, t) + {\bar {J}} ({\mathbf{p}}', {\mathbf{p}}, {\mathbf{r}}, t)]
\}
$
for ${\mathbf{p}} = \hbar   {\mathbf{k}}   $,
the correlation becomes
\begin{eqnarray}
{\cal{S}}(t, t') 
&=& 
(g \mu_{\rm{B}})^2  L^4     V   \delta (t-t')
\int  \frac{{d {\mathbf{k}}}}{{(2\pi)^3}}   \int  \frac{{d {\mathbf{k}}''}}{{(2\pi)^3}}   (v_{k_x})^2  ({\tau _{\mathbf{k}}})^2      \nonumber   \\
&\times & \langle  \langle  \delta ({\mathbf{r}} - {\mathbf{r}}') [{\bar {J}} ({\mathbf{p}}'', {\mathbf{p}}) + {\bar {J}} ({\mathbf{p}}, {\mathbf{p}}'')]   \rangle \rangle
   \nonumber    \\
&-& 
(g \mu_{\rm{B}})^2  L^4     V   \delta (t-t')
\int  \frac{{d {\mathbf{k}}}}{{(2\pi)^3}}   \int  \frac{{d {\mathbf{k}}'}}{{(2\pi)^3}}  v_{k_x} v_{k'_x} {\tau _{\mathbf{k}}}   {\tau _{{\mathbf{k}}'}}       \nonumber   \\
&\times & \langle  \langle \delta ({\mathbf{r}} - {\mathbf{r}}') [{\bar {J}} ({\mathbf{p}}, {\mathbf{p}}') + {\bar {J}} ({\mathbf{p}}', {\mathbf{p}})]   \rangle \rangle ,
 \label{eqn:SMEq2}
\end{eqnarray}
where
$  {\bar {J}}({\mathbf{p}}, {\mathbf{p}}') =  W({\mathbf{p}}, {\mathbf{p}}')  {\bar {f}}_{{\mathbf{p}}} (1+{\bar {f}}_{{\mathbf{p}}'})   $
and
$    W({\mathbf{p}}, {\mathbf{p}}') =  - ({2}/{\hbar }) \mid V_{{\mathbf{p}}, {\mathbf{p}}'} \mid ^2
  {\rm{Im}} G^{\rm{r}}_{{\mathbf{p}}', \omega , {\mathbf{r}}, t} \mid _{\omega =\omega _{\mathbf{p}}}  $;
the retarded Green's function becomes
$  {\rm{Im}}  G^{\rm{r}} = - \pi \delta (\hbar \omega - \hbar  \omega _{\mathbf{p}})  $
within the quasiparticle approximation.
Since the Fourier component becomes momentum-independent \cite{haug}
under the assumption that the impurity potential is localized in space,
finally,
we obtain the correlation in the main text; 
\begin{eqnarray}
 {\cal{S}}(t, t')
= 2 (g \mu_{\rm{B}})^2  L  \delta (t-t')   \int  \frac{d {\mathbf{k}}}{(2\pi)^3}    {\tau _{\mathbf{k}}}  (v_{k_x})^2  
     \langle {\bar {f}}_{\mathbf{k}}  (1 + {\bar {f}}_{\mathbf{k}})  \rangle.        \       \      \       
\label{eqn:SMEq3}
\end{eqnarray}

\subsection{Spatial average in diffusive regime}
\label{subsec:SM1-2}


Deriving the effective kinetic equation of magnons in the diffusive regime,
we evaluate the spatial average $   \langle {\bar {f}}_{\mathbf{k}}(x)  [1 + {\bar {f}}_{\mathbf{k}}(x)] \rangle $
in Eq. (\ref{eqn:SMEq3}).
Assuming a steady state in terms of time and multiplying
the semiclassical Boltzmann equation by $\partial _x$,
it becomes 
$  \partial _x^2  {\bar {f}}_{{\mathbf{k}}}(x) = -(1/l_{{\mathbf{k}}}) \partial _x  {\bar {f}}_{{\mathbf{k}}}(x)    $,
where $  l_{{\mathbf{k}}} :=  v_{k_x}  \tau _{\mathbf{k}} \not=0 $ is the mean free path \cite{NoiseRev} of magnons.
The solution under the boundary conditions, 
$  {\bar {f}}_{{\mathbf{k}}}(x=-L/2) \equiv  f_{\rm{L}}    $ and $  {\bar {f}}_{{\mathbf{k}}}(x= L/2) \equiv  f_{\rm{R}}    $,
is given by
$ {\bar {f}}_{{\mathbf{k}}}(x) =  [(f_{\rm{L}}-f_{\rm{R}})/2{\rm{sinh}}(L/2l_{{\mathbf{k}}})] {\rm{e}}^{-x/ l_{{\mathbf{k}}}}  
+ (f_{\rm{L}}+f_{\rm{R}})/2 - [(f_{\rm{L}}-f_{\rm{R}})/2] {\rm{coth}}(L/2l_{{\mathbf{k}}})
$,
and therefore the semiclassical Boltzmann equation can be rewritten as
$  \partial _x^2  {\bar {f}}_{{\mathbf{k}}}(x) = (1/l_{{\mathbf{k}}}^2)  
 [(f_{\rm{L}}-f_{\rm{R}})/2{\rm{sinh}}(L/2l_{{\mathbf{k}}})] {\rm{e}}^{-x/ l_{{\mathbf{k}}}}  
$;
in the diffusive regime \cite{NoiseRev} $ l_{{\mathbf{k}}}  \ll  L $, it reduces to \cite{Nagaev1/3}
 $\partial _x^2  {\bar {f}}_{{\mathbf{k}}}(x)   =  0$.
This is identified with the effective kinetic equation of magnons in the diffusive region.
The solution under the boundary conditions becomes
$  {\bar {f}}_{{\mathbf{k}}}(x) =  (f_{\rm{R}}-f_{\rm{L}})(x/L) +(f_{\rm{L}}+f_{\rm{R}})/2   $,
and 
a straightforward calculation provides (see the next section for details)
\begin{eqnarray}
  \langle {\bar {f}}_{\mathbf{k}}(x)  [1 + {\bar {f}}_{\mathbf{k}}(x)] \rangle
&=& [(f_{\rm{R}}-f_{\rm{L}})/2] {\rm{coth}}(\beta _{\rm{L}} \hbar \omega _{\mathbf{k}}/2 - \beta _{\rm{R}} \hbar \omega _{\mathbf{k}}/2)   \nonumber   \\
&+& (1/3) (f_{\rm{L}}-f_{\rm{R}})^2,
\label{eqn:SMBBB}
\end{eqnarray}
where
$ \beta _{\rm{L(R)}}=  (k_{\rm{B}} T_{\rm{L(R)}})^{-1}$.
Since (see the next section for details)
\begin{eqnarray}
 (f_{\rm{L}}-f_{\rm{R}})^2 
&=&  - (f_{\rm{R}}-f_{\rm{L}}) {\rm{coth}}(\beta _{\rm{L}} \hbar \omega _{\mathbf{k}}/2 - \beta _{\rm{R}} \hbar \omega _{\mathbf{k}}/2)     \nonumber   \\
&-& k_{\rm{B}} T_{\rm{L}} \partial  f_{\rm{L}}(\epsilon )/\partial \epsilon 
- k_{\rm{B}} T_{\rm{R}} \partial  f_{\rm{R}}(\epsilon )/\partial \epsilon 
\label{eqn:SMCCC}
\end{eqnarray}
with
$  f_{\rm{L(R)}}(\epsilon ) := ({\rm{e}}^{\beta _{\rm{L(R)}} \epsilon }-1)^{-1} $
and
$\epsilon :=  \hbar \omega _{\mathbf{k}} $,
finally, we obtain the spatial average in the main text;
\begin{eqnarray}
    \langle {\bar {f}}_{\mathbf{k}}(x)  [1 + {\bar {f}}_{\mathbf{k}}(x)] \rangle
&=& \frac{f_{\rm{R}}-f_{\rm{L}}}{6} 
{\rm{coth}} \Big(\frac{\beta _{\rm{L}} \hbar \omega _{\mathbf{k}}}{2} 
- \frac{\beta _{\rm{R}} \hbar \omega _{\mathbf{k}}}{2} \Big)      \nonumber    \\
&-& \frac{1}{3}  \Big( k_{\rm{B}} T_{\rm{L}} \frac{\partial  f_{\rm{L}}(\epsilon )}{\partial \epsilon} 
+ k_{\rm{B}} T_{\rm{R}} \frac{\partial  f_{\rm{R}}(\epsilon )}{\partial \epsilon} \Big).          \        \         \       \        \       \       \     
  \label{eqn:SMAAA}            
\end{eqnarray}

\subsection{Details of the calculation}
\label{subsec:SM1-2-2}

We provide details on the derivation of Eqs. (\ref{eqn:SMBBB})-(\ref{eqn:SMAAA})
in this Appendix;
using the effective kinetic equation of magnons in the diffusive region,
$  \partial _x^2  {\bar {f}}_{{\mathbf{k}}}(x)   =  0  $,
we evaluate the spatial average 
$ \langle {\bar {f}}_{\mathbf{k}}  (1 + {\bar {f}}_{\mathbf{k}})  \rangle  
:= (1/L) \int_{-L/2}^{L/2} dx {\bar {f}}_{\mathbf{k}}  (1 + {\bar {f}}_{\mathbf{k}}) $
under the boundary conditions 
$  {\bar {f}}_{{\mathbf{k}}}(x=-L/2) \equiv  f_{\rm{L}}    $ and $  {\bar {f}}_{{\mathbf{k}}}(x= L/2) \equiv  f_{\rm{R}}    $,
where 
the temperature is $T_{\rm{L}}$ and $T_{\rm{R}}$, respectively.

First, since the solution is given by 
$  {\bar {f}}_{{\mathbf{k}}}(x) =  (f_{\rm{R}}-f_{\rm{L}})(x/L) +(f_{\rm{L}}+f_{\rm{R}})/2   $,
it becomes
\begin{eqnarray}
 \langle {\bar {f}}_{\mathbf{k}}  (1 + {\bar {f}}_{\mathbf{k}})  \rangle  
 = \frac{1}{2} \big[(f_{\rm{L}}+f_{\rm{R}}) + \frac{2}{3}(f_{\rm{L}}^2+f_{\rm{R}}^2+f_{\rm{L}}f_{\rm{R}}) \big].        \        \       \    
\label{eqn:SMEq4}
\end{eqnarray}

Second,
we rewrite $ (f_{\rm{L}}+f_{\rm{R}}) $ in Eq. (\ref{eqn:SMEq4}).
Defining
$  f_{\rm{L(R)}}(\epsilon ) := ({\rm{e}}^{\beta _{\rm{L(R)}} \epsilon }-1)^{-1} $
and
$\epsilon :=  \hbar \omega _{\mathbf{k}} $,
it becomes
\begin{eqnarray}
 f_{\rm{L(R)}}(\epsilon ) = \frac{{\rm{coth}}(\beta _{\rm{L(R)}} \epsilon/2)-1}{2}.
\label{eqn:SMEq5}
\end{eqnarray}
Using the relation
\begin{eqnarray}
{\rm{coth}}(\beta _{\rm{L}} \epsilon) - {\rm{coth}}(\beta _{\rm{R}} \epsilon)
= - \frac{ {\rm{sinh}}(\beta _{\rm{L}} \epsilon - \beta _{\rm{R}} \epsilon)  }
                    { {\rm{sinh}}(\beta _{\rm{L}} \epsilon){\rm{sinh}}(\beta _{\rm{R}} \epsilon)  },
 \label{eqn:SMEq6}  
\end{eqnarray}
a straightforward calculation provides
\begin{subequations}
\begin{eqnarray}
{\rm{coth}}(\beta _{\rm{L}} \epsilon)  {\rm{coth}}(\beta _{\rm{R}} \epsilon) -1         
\        \        \       \        \        \         \        \         \         \            \             \             \           \         \        \         \         \            \             \             \                \nonumber      \\
=    \frac{ {\rm{sinh}}(\beta _{\rm{L}} \epsilon - \beta _{\rm{R}} \epsilon )   }{ {\rm{sinh}}(\beta _{\rm{L}} \epsilon) {\rm{sinh}}(\beta _{\rm{R}} \epsilon)   }
\frac{ {\rm{cosh}}(\beta _{\rm{L}} \epsilon - \beta _{\rm{R}} \epsilon )   }{{\rm{sinh}}(\beta _{\rm{L}} \epsilon - \beta _{\rm{R}} \epsilon )    }
  \        \         \         \           \     
\label{eqn:SMEq7a}   \\
=      [{\rm{coth}}(\beta _{\rm{R}} \epsilon) - {\rm{coth}}(\beta _{\rm{L}} \epsilon)]    
\frac{ {\rm{cosh}}(\beta _{\rm{L}} \epsilon - \beta _{\rm{R}} \epsilon )   }{{\rm{sinh}}(\beta _{\rm{L}} \epsilon - \beta _{\rm{R}} \epsilon ) }
\label{eqn:SMEq7b}    \\
=      [{\rm{coth}}(\beta _{\rm{R}} \epsilon) - {\rm{coth}}(\beta _{\rm{L}} \epsilon)]    
{\rm{coth}}( \beta _{\rm{L}} \epsilon - \beta _{\rm{R}} \epsilon ),
\label{eqn:SMEq7c}
\end{eqnarray}
\end{subequations}
and thereby
\begin{eqnarray}
{\rm{coth}}(\beta _{\rm{L}} \epsilon/2)  {\rm{coth}}(\beta _{\rm{R}} \epsilon/2) -1
&=&   [{\rm{coth}}(\beta _{\rm{R}} \epsilon/2) - {\rm{coth}}(\beta _{\rm{L}} \epsilon/2)]      \nonumber      \\
&\times & {\rm{coth}}( \beta _{\rm{L}} \epsilon/2 - \beta _{\rm{R}} \epsilon/2 ).
\label{eqn:SMEq8}
\end{eqnarray}
From Eq. (\ref{eqn:SMEq5}) we obtain 
\begin{eqnarray}
 {\rm{coth}}(\beta _{\rm{L(R)}} \epsilon/2) =  1+ 2 f_{\rm{L(R)}}(\epsilon ),
\label{eqn:SMEq9}
\end{eqnarray}
and see that
\begin{eqnarray}
 {\rm{coth}}(\beta _{\rm{R}} \epsilon/2) -  {\rm{coth}}(\beta _{\rm{L}} \epsilon/2) =  2 [f_{\rm{R}}(\epsilon ) - f_{\rm{L}}(\epsilon )].
\label{eqn:SMEq10}
\end{eqnarray}
On top of it, from Eq. (\ref{eqn:SMEq9}) we obtain
\begin{subequations}
\begin{eqnarray}
 {\rm{coth}}(\beta _{\rm{L}} \epsilon/2)  {\rm{coth}}(\beta _{\rm{R}} \epsilon/2) 
&=&  [1+ 2 f_{\rm{L}}(\epsilon )]  [1+ 2 f_{\rm{R}}(\epsilon )]       \           \          \           \           \         \           \            \             \            \               \\
\label{eqn:SMEq11a}
&=& 1+ 2 [f_{\rm{L}}(\epsilon ) + f_{\rm{R}}(\epsilon )]     \nonumber    \\
&+& 4 f_{\rm{L}}(\epsilon )  f_{\rm{R}}(\epsilon ),
\label{eqn:SMEq11b}
\end{eqnarray}
\end{subequations}
and see that 
\begin{eqnarray}
 {\rm{coth}}(\beta _{\rm{L}} \epsilon/2)  {\rm{coth}}(\beta _{\rm{R}} \epsilon/2) -1
&=& 2 f_{\rm{L}}(\epsilon ) [1+f_{\rm{R}}(\epsilon )]     \nonumber   \\
&+& 2 f_{\rm{R}}(\epsilon ) [1+f_{\rm{L}}(\epsilon )].       \           \          \           \            \              \             \               \             \           
\label{eqn:SMEq12}
\end{eqnarray}
Putting Eqs. (\ref{eqn:SMEq10}) and (\ref{eqn:SMEq12}) into Eq. (\ref{eqn:SMEq8}),
finally, $ (f_{\rm{L}}+f_{\rm{R}}) $ is rewritten by 
\begin{eqnarray}
f_{\rm{L}}(\epsilon ) + f_{\rm{R}}(\epsilon )
&=& [f_{\rm{R}}(\epsilon ) - f_{\rm{L}}(\epsilon )]
{\rm{coth}}(\beta _{\rm{L}} \epsilon/2 - \beta _{\rm{R}} \epsilon/2)           \nonumber    \\
&-& 2 f_{\rm{L}}(\epsilon )  f_{\rm{R}}(\epsilon ),
\label{eqn:SMEq13}
\end{eqnarray}
and thus Eq. (\ref{eqn:SMEq4}) becomes Eq. (\ref{eqn:SMBBB});
\begin{eqnarray}
 \langle {\bar {f}}_{\mathbf{k}}  (1 + {\bar {f}}_{\mathbf{k}})  \rangle  
 &=& \frac{1}{2} \big[
(f_{\rm{R}} - f_{\rm{L}})
{\rm{coth}}(\beta _{\rm{L}} \epsilon/2 - \beta _{\rm{R}} \epsilon/2)       \nonumber    \\
& +& \frac{2}{3}(f_{\rm{L}} - f_{\rm{R}})^2 \big].
\label{eqn:SMEq14}
\end{eqnarray}

Next, we rewrite $ (f_{\rm{L}} - f_{\rm{R}})^2  $ in Eq. (\ref{eqn:SMEq14}).
Since
\begin{subequations}
\begin{eqnarray}
\partial  f_{\rm{L(R)}}(\epsilon )/\partial \epsilon 
&=& - \beta _{\rm{L(R)}}  {\rm{e}}^{\beta _{\rm{L(R)}} \epsilon }/({\rm{e}}^{\beta _{\rm{L(R)}} \epsilon }-1)^2,     \       \       \       \        \          \        \          \       \ 
    \label{eqn:SMEq15a}    \\
 f_{\rm{L(R)}}(1+ f_{\rm{L(R)}})
&=&
 {\rm{e}}^{\beta _{\rm{L(R)}} \epsilon }/({\rm{e}}^{\beta _{\rm{L(R)}} \epsilon }-1)^2,      \       \       \       \        \          \        \          \       \ 
\label{eqn:SMEq15b}
\end{eqnarray}
\end{subequations}
we obtain
\begin{eqnarray}
f_{\rm{L(R)}}(1+ f_{\rm{L(R)}}) = - (1/\beta _{\rm{L(R)}})     \partial  f_{\rm{L(R)}}(\epsilon )/\partial \epsilon ,
    \label{eqn:SMEq16}    
\end{eqnarray}
and see that 
\begin{eqnarray}
f_{\rm{L}}(1+ f_{\rm{L}}) + f_{\rm{R}}(1+ f_{\rm{R}})
 &=& - (1/\beta _{\rm{L}})     \partial  f_{\rm{L}}(\epsilon )/\partial \epsilon         \nonumber     \\
& -& (1/\beta _{\rm{R}})     \partial  f_{\rm{R}}(\epsilon )/\partial \epsilon .        \         \       \      \        \         \      
    \label{eqn:SMEq17}    
\end{eqnarray}
Putting Eqs. (\ref{eqn:SMEq10}) and (\ref{eqn:SMEq12}) into Eq. (\ref{eqn:SMEq8}),
we remind that 
\begin{eqnarray}
f_{\rm{L}}(1+ f_{\rm{R}}) + f_{\rm{R}}(1+ f_{\rm{L}})
= (f_{\rm{R}} - f_{\rm{L}})
{\rm{coth}}(\beta _{\rm{L}} \epsilon/2 - \beta _{\rm{R}} \epsilon/2),      \         \         \          \           \    
\label{eqn:SMEq18}
\end{eqnarray}
and see that $[{\rm{Eq.}} (\ref{eqn:SMEq17})] - [{\rm{Eq.}} (\ref{eqn:SMEq18})] = (f_{\rm{L}} - f_{\rm{R}})^2 $.
Thus $ (f_{\rm{L}}-f_{\rm{R}})^2 $ is rewritten by Eq. (\ref{eqn:SMCCC});
\begin{eqnarray}
  (f_{\rm{L}}-f_{\rm{R}})^2 &=&  - (f_{\rm{R}}-f_{\rm{L}}) {\rm{coth}}(\beta _{\rm{L}} \hbar \omega _{\mathbf{k}}/2 - \beta _{\rm{R}} \hbar \omega _{\mathbf{k}}/2)       \nonumber    \\
&-& k_{\rm{B}} T_{\rm{L}} \partial  f_{\rm{L}}(\epsilon )/\partial \epsilon 
- k_{\rm{B}} T_{\rm{R}} \partial  f_{\rm{R}}(\epsilon )/\partial \epsilon .       \            \         \           \           \          \   
\label{eqn:SMEq19}
\end{eqnarray}

Finally,
putting Eq. (\ref{eqn:SMEq19}) into Eq. (\ref{eqn:SMEq14}),
we obtain the spatial average [Eq. (\ref{eqn:SMAAA})] in the main text;
\begin{eqnarray}
    \langle {\bar {f}}_{\mathbf{k}}  (1 + {\bar {f}}_{\mathbf{k}}) \rangle
&=& \frac{f_{\rm{R}}-f_{\rm{L}}}{6} 
{\rm{coth}} \Big(\frac{\beta _{\rm{L}} \hbar \omega _{\mathbf{k}}}{2} 
- \frac{\beta _{\rm{R}} \hbar \omega _{\mathbf{k}}}{2} \Big)        \nonumber     \\
&-& \frac{1}{3}  \Big( k_{\rm{B}} T_{\rm{L}} \frac{\partial  f_{\rm{L}}}{\partial \epsilon} 
+ k_{\rm{B}} T_{\rm{R}} \frac{\partial  f_{\rm{R}}}{\partial \epsilon} \Big).   
  \label{eqn:Eq21}    
\end{eqnarray}

\section{Quantum kinetic equation}
\label{sec:SM2}

In this Appendix starting from the quantum kinetic equation, we provide details on the derivation of the Boltzmann equation for magnons [Eq. (\ref{eqn:Eq1})].
Defining $ h := \hbar  \omega  -  \hbar \omega _{\mathbf{k}} $,
the quantum kinetic equation is given by 
\begin{eqnarray}
[h, G^{<}]_{{\mathbf{k}}, \omega , {\mathbf{r}}, t} 
= (G^{<}_{{\mathbf{k}}, \omega , {\mathbf{r}}, t} \Sigma ^{>}_{{\mathbf{k}}, \omega , {\mathbf{r}}, t}
- G^{>}_{{\mathbf{k}}, \omega , {\mathbf{r}}, t} \Sigma ^{<}_{{\mathbf{k}}, \omega , {\mathbf{r}}, t}),    \     \     \    \   
 \label{eqn:AppC_Eq1} 
\end{eqnarray}
where 
\begin{eqnarray}
[h, G^{<}]_{{\mathbf{k}}, \omega , {\mathbf{r}}, t} 
&:=& -i \Big(
\frac{\partial h}{\partial  t} \frac{\partial G^{<}_{{\mathbf{k}}, \omega , {\mathbf{r}}, t}}{\partial  \omega }
-\frac{\partial h}{\partial  \omega } \frac{\partial G^{<}_{{\mathbf{k}}, \omega , {\mathbf{r}}, t}}{\partial  t}    \nonumber    \\
&-&\frac{\partial h}{\partial  {\mathbf{r}} } \frac{\partial G^{<}_{{\mathbf{k}}, \omega , {\mathbf{r}}, t}}{\partial  {\mathbf{k}}}
+\frac{\partial h}{\partial  {\mathbf{k}} } \frac{\partial G^{<}_{{\mathbf{k}}, \omega , {\mathbf{r}}, t}}{\partial  {\mathbf{r}}}
\Big).
 \label{eqn:AppC_Eq2} 
\end{eqnarray}
Since $h$ of our system is time-independent and spatially uniform,
\begin{eqnarray}
[h, G^{<}]_{{\mathbf{k}}, \omega , {\mathbf{r}}, t} 
= i \hbar  \Big(\frac{\partial}{\partial t} + {\mathbf{v}} \cdot   \frac{\partial}{\partial   {\mathbf{r}}} \Big)   G^{<}_{{\mathbf{k}}, \omega , {\mathbf{r}}, t}
 \label{eqn:AppC_Eq2} 
\end{eqnarray}
and the quantum kinetic equation becomes 
\begin{eqnarray}
 (\partial _t + {\mathbf{v}} \cdot  \partial _{{\mathbf{r}}}) G^{<}_{{\mathbf{k}}, \omega , {\mathbf{r}}, t}
& =& \frac{1}{{i\hbar }} (G^{<}_{{\mathbf{k}}, \omega , {\mathbf{r}}, t} \Sigma ^{>}_{{\mathbf{k}}, \omega , {\mathbf{r}}, t}    \nonumber    \\
&-& G^{>}_{{\mathbf{k}}, \omega , {\mathbf{r}}, t} \Sigma ^{<}_{{\mathbf{k}}, \omega , {\mathbf{r}}, t}).
 \label{eqn:AppC_Eq3} 
\end{eqnarray}
Within the Born approximation \cite{haug}, the self-energy due to magnon-dilute impurity scattering is given by
\begin{eqnarray}
\Sigma _{{\mathbf{k}}, \omega , {\mathbf{r}}, t} = \frac{{V}}{{(2\pi)^3}} \int d {\mathbf{k}}' 
G_{{\mathbf{k}}', \omega , {\mathbf{r}}, t} \mid    V_{{\mathbf{k}}, {\mathbf{k}}'}   \mid ^2,
 \label{eqn:AppC_Eq4} 
\end{eqnarray}
and thereby the quantum kinetic equation of magnons becomes
\begin{eqnarray}
 (\partial _t + {\mathbf{v}} \cdot  \partial _{{\mathbf{r}}}) G^{<}_{{\mathbf{p}}, \omega , {\mathbf{r}}, t}
 &=& \frac{1}{{i\hbar }}   \frac{{V}}{{(2\pi \hbar )^3}}  \int d {\mathbf{p}}'  \mid    V_{{\mathbf{p}}, {\mathbf{p}}'}   \mid ^2     \label{eqn:AppC_Eq5}  \\
&\cdot & (G^{<}_{{\mathbf{p}}, \omega , {\mathbf{r}}, t} G^{>}_{{\mathbf{p}}', \omega , {\mathbf{r}}, t}
- G^{>}_{{\mathbf{p}}, \omega , {\mathbf{r}}, t} G ^{<}_{{\mathbf{p}}', \omega , {\mathbf{r}}, t}).   \nonumber 
\end{eqnarray}
Following the straightforward calculation procedure explained in detail in the main text, 
it is easily seen that Eq. (\ref{eqn:AppC_Eq5}) reduces to the Boltzmann equation for magnons [Eq. (\ref{eqn:Eq1})];
the Kadanoff-Baym ansatz ensures
$ G^{<}_{{\mathbf{p}}, \omega , {\mathbf{r}}, t} 
= 2 i {\rm{Im}}  G^{\rm{r}}_{{\mathbf{p}}, \omega , {\mathbf{r}}, t} f_{{\mathbf{p}}, \omega , {\mathbf{r}}, t} $ and
$ G^{>}_{{\mathbf{p}}, \omega , {\mathbf{r}}, t} 
= 2 i {\rm{Im}}  G^{\rm{r}}_{{\mathbf{p}}, \omega , {\mathbf{r}}, t} (1+f_{{\mathbf{p}}, \omega , {\mathbf{r}}, t}) $.
Thereby Eq. (\ref{eqn:AppC_Eq5}) becomes
\begin{eqnarray}
 (\partial _t + {\mathbf{v}} \cdot  \partial _{{\mathbf{r}}}) 
{\rm{Im}} G^{\rm{r}}_{{\mathbf{p}}, \omega , {\mathbf{r}}, t}   f_{{\mathbf{p}}}
 &=& \frac{2}{{\hbar }}   \frac{{V}}{{(2\pi \hbar )^3}}  \int d {\mathbf{p}}'  \mid    V_{{\mathbf{p}}, {\mathbf{p}}'}   \mid ^2    \     \      \     \     \        \
 \label{eqn:AppC_Eq6}  \\
&\cdot &     {\rm{Im}} G^{\rm{r}}_{{\mathbf{p}}, \omega , {\mathbf{r}}, t}  {\rm{Im}} G^{\rm{r}}_{{\mathbf{p}}', \omega , {\mathbf{r}}, t}   \nonumber    \\
&\cdot & [ f_{{\mathbf{p}}}(1+ f_{{\mathbf{p}}'}) -   f_{{\mathbf{p}}'}(1+ f_{{\mathbf{p}}}) ].   \nonumber 
\end{eqnarray}
Within the quasiparticle approximation, we see that $  {\rm{Im}}  G^{\rm{r}} = - \pi \delta (\hbar \omega - \hbar  \omega _{\mathbf{p}})  $.
Therefore integrating Eq. (\ref{eqn:AppC_Eq6}) over the frequency $\omega $, $ \int d \omega $, 
it becomes
\begin{eqnarray}
 (\partial _t + {\mathbf{v}} \cdot  \partial _{{\mathbf{r}}})    f_{{\mathbf{p}}}
 &=& \frac{2}{{\hbar }}   \frac{{V}}{{(2\pi \hbar )^3}}  \int d {\mathbf{p}}'  \mid    V_{{\mathbf{p}}, {\mathbf{p}}'}   \mid ^2    \     \      \     \     \        \
 \label{eqn:AppC_Eq7}  \\
&\cdot &  {\rm{Im}} G^{\rm{r}}_{{\mathbf{p}}', \omega , {\mathbf{r}}, t}\mid _{\omega =\omega _{{\mathbf{p}}}}   \nonumber    \\
&\cdot & [ f_{{\mathbf{p}}}(1+ f_{{\mathbf{p}}'}) -   f_{{\mathbf{p}}'}(1+ f_{{\mathbf{p}}}) ].  \nonumber 
\end{eqnarray}
We thus reach the Boltzmann equation for magnons [Eq. (\ref{eqn:Eq1})].

\section{Langevin source}
\label{sec:SM3}

In this Appendix, we add an explanation about the derivation of the correlations between the Langevin sources.
The deviation of a Poisson distribution is equal to the square root of the mean value,
and the Poissonian statistics generally apply to random processes \cite{TextbookQuantumOptics}. 
In this work, following Refs. [\onlinecite{Kogan,NoiseRev}] we assume a Poisson process to describe the randomness of the scattering process
and that the particle currents $J$ between the states are independent elementary processes; 
the currents are correlated only when they describe the same process, e.g.,  identical initial and final states, space point, and time. 
Those result in 
\begin{eqnarray}
\langle \delta J({\mathbf{p}}_1, {\mathbf{p}}_2, {\mathbf{r}}, t)  
\delta J({\mathbf{p}}'_1, {\mathbf{p}}'_2, {\mathbf{r}}', t') \rangle 
&=&   \frac{(2 \pi \hbar)^6}{V}  \delta ({\mathbf{p}}_1 - {\mathbf{p}}'_1)   \nonumber    \\
&\cdot & \delta ({\mathbf{p}}_2 - {\mathbf{p}}'_2)      \nonumber    \\
&\cdot & \delta ({\mathbf{r}} - {\mathbf{r}}')  \delta (t-t')     \nonumber     \\
&\cdot & {\bar {J}} ({\mathbf{p}}_1, {\mathbf{p}}_2, {\mathbf{r}}, t).
 \label{eqn:AppD_Eq1} 
\end{eqnarray}
Using Eqs. (\ref{eqn:Eq4}) and (\ref{eqn:AppD_Eq1}), we obtain the correlations between the Langevin sources;
\begin{eqnarray}
\langle  \xi ({\mathbf{r}}, {\mathbf{p}}, t)   \xi ({\mathbf{r}}', {\mathbf{p}}', t')    \rangle   
&=& V \delta ({\mathbf{r}} - {\mathbf{r}}')  \delta (t-t')
\{ 
\delta ({\mathbf{p}} - {\mathbf{p}}')  \int   d {\mathbf{p}}''       \nonumber      \\
&\cdot &[{\bar {J}} ({\mathbf{p}}'', {\mathbf{p}}, {\mathbf{r}}, t) + {\bar {J}} ({\mathbf{p}}, {\mathbf{p}}'', {\mathbf{r}}, t)]     \nonumber    \\
&-& [{\bar {J}} ({\mathbf{p}}, {\mathbf{p}}', {\mathbf{r}}, t) + {\bar {J}} ({\mathbf{p}}', {\mathbf{p}}, {\mathbf{r}}, t)]
\}.
 \label{eqn:AppD_Eq2} 
\end{eqnarray}

For more detail about Poisson statistics, see Appendix A of Ref. [\onlinecite{TextbookQuantumOptics}] if necessary.

\section{Relaxation time}
\label{sec:SM4}

In this Appendix, we provide details on the derivation of the relaxation time $ {\tau _{\mathbf{k}}}$.
Assuming a steady state in terms of time, the Boltzmann equation becomes
\begin{eqnarray}
  {\mathbf{v}}_{\mathbf{k}} \cdot  \partial _{{\mathbf{r}}} {\bar {f}}_{{\mathbf{k}}}   
= - \frac{\bar {f}_{\mathbf{k}} - f_{\rm{eq.}}}{\tau _{\mathbf{k}}},   
 \label{eqn:AppBEq1} 
\end{eqnarray}
where
$ {\mathbf{v}}_{\mathbf{k}} = \partial _{{\mathbf{k}}} \omega _{\mathbf{k}}$ is the magnon velocity.
Defining $  g_{\mathbf{k}} := \bar {f}_{\mathbf{k}} - f_{\rm{eq.}}$, the Boltzmann equation is rewritten as
\begin{subequations}
\begin{eqnarray}
  g_{\mathbf{k}} &=& - \tau _{\mathbf{k}}  {\mathbf{v}}_{\mathbf{k}}  \cdot  {\mathbf{\nabla}} T    \frac{ \partial  {\bar {f}}_{{\mathbf{k}}}}{\partial T},   \\
 \label{eqn:AppBEq2} 
 &=& - \tau _{\mathbf{k}}  {\mathbf{v}}_{\mathbf{k}}  \cdot  {\mathbf{\nabla}} T  
            \frac{ \partial    (g_{\mathbf{k}} + f_{\rm{eq.}})}{\partial T},
\label{eqn:AppBEq3} 
\end{eqnarray}
\end{subequations}
where we have used the relation $    \partial _{{\mathbf{r}}} =  {\mathbf{\nabla}} T \partial /\partial T $
and $   \bar {f}_{\mathbf{k}} = g_{\mathbf{k}} +  f_{\rm{eq.}}$.
Substituting $  g_{\mathbf{k}}  $ into Eq. (\ref{eqn:AppBEq3}) sequentially, we obtain
\begin{eqnarray}
  g_{\mathbf{k}} = - \tau _{\mathbf{k}}  {\mathbf{v}}_{\mathbf{k}}  \cdot  {\mathbf{\nabla}} T  
            \frac{ \partial    f_{\rm{eq.}}}{\partial T} + {\cal{O}} \big(( {\mathbf{\nabla}} T  )^2\big).
\label{eqn:AppBEq4} 
\end{eqnarray}
On the other hand, from the collision integral, we obtain
\begin{eqnarray}
  \frac{g_{\mathbf{k}}}{{\tau _{\mathbf{k}}}} =
\Sigma _{{\mathbf{k}}'} W({\mathbf{k}}, {\mathbf{k}}') (g_{{\mathbf{k}}} - g_{{\mathbf{k}}'} ).
\label{eqn:AppBEq5}
\end{eqnarray}
Putting Eq. (\ref{eqn:AppBEq4}) into Eq. (\ref{eqn:AppBEq5}), it is seen that
\begin{eqnarray}
\frac{{\mathbf{v}}_{\mathbf{k}}}{{\tau _{\mathbf{k}}}} = \Sigma _{{\mathbf{k}}'} W({\mathbf{k}}, {\mathbf{k}}')
({\mathbf{v}}_{\mathbf{k}} - {\mathbf{v}}_{{\mathbf{k}}'}).
\label{eqn:AppBEq6}
\end{eqnarray}
Multiplying $ {\mathbf{v}}_{\mathbf{k}}$ and divided by $ \mid  {\mathbf{v}}_{\mathbf{k}}  \mid ^2$, finally
the relaxation time $ {\tau _{\mathbf{k}}}$ is given by 
\begin{eqnarray}
\frac{ 1}{{\tau _{\mathbf{k}}}} = \Sigma _{{\mathbf{k}}'} W({\mathbf{k}}, {\mathbf{k}}')
\Big(1- \frac{{\mathbf{v}}_{\mathbf{k}}\cdot {\mathbf{v}}_{{\mathbf{k}}'}}{\mid  {\mathbf{v}}_{\mathbf{k}}  \mid ^2}\Big).
\label{eqn:AppBEq7}
\end{eqnarray}

\bibliography{PumpingRef}

\begin{thebibliography}{47}
\bibitem[{Whi()}]{WhiteNoise}
\bibinfo{note}{Since we assume a Gaussian distribution for the Langevin source,
  the noise spectrum becomes white noise process.}
\end{thebibliography}

\end{document}